\documentclass[twoside]{ae100prg}
\usepackage{graphicx}
\usepackage[breaklinks]{hyperref}
\usepackage{booktabs}

\newcommand{\dd}{{\rm{d}}} 
\newcommand{\rovno}{\!\!\!\!& = &\!\!\!\!} 
\newcommand{\ssqrt}{{\textstyle\frac1{\sqrt{2}}}}          

\newcommand{\boldu}{\mbox{\boldmath$u$}}                   
\newcommand{\bolde}{\mbox{\boldmath$e$}}                   
\newcommand{\boldk}{\mbox{\boldmath$k$}}                   
\newcommand{\boldl}{\mbox{\boldmath$l$}}                   
\newcommand{\boldm}{\mbox{\boldmath$m$}}                   
\newcommand{\boldZ}{\mbox{\boldmath$Z$}}                   

\begin{document}

\title{Geodesic deviation in Kundt spacetimes of any dimension}

\author{Robert \v{S}varc and Ji\v{r}\'{\i} Podolsk\'{y}}

\address{Institute of Theoretical Physics, Charles University in Prague, V~Hole\v{s}ovi\v{c}k\'{a}ch 2, 180~00 Praha 8, Czech Republic}

\email{robert.svarc@mff.cuni.cz podolsky@mbox.troja.mff.cuni.cz}

\begin{abstract}
Using the invariant form of the equation of geodesic deviation, which
describes relative motion of free test particles, we investigate
a general family of $D$-dimensional Kundt spacetimes.
We demonstrate that local influence of the gravitational field can be
naturally decomposed into Newton-type tidal effects typical for
type~II spacetimes, longitudinal deformations mainly present
in spacetimes of algebraic type~III, and type~N purely transverse
effects corresponding to gravitational waves with ${\frac{1}{2}D(D-3)}$
independent polarization states. We explicitly study the most
important examples, namely exact pp-waves, gyratons, and VSI spacetimes.
This analysis helps us to clarify the geometrical and physical
interpretation of the Kundt class of nonexpanding, nontwisting
and shearfree geometries.
\end{abstract}

\section{Geometry of Kundt spacetimes}

The scalars $\Theta$ (expansion), $A^2$ (twist) and $\sigma^2$ (shear) characterizing optical properties of an affinely parameterized geodesic null congruence $k^a$ are
\begin{equation}
\Theta=\frac{1}{D-2}k^{a}_{\; ;a}  \ , \quad A^2=-k_{[a;b]}k^{a;b} \ , \quad \sigma^2=k_{(a;b)}k^{a;b}-\frac{1}{D-2}(k^{a}_{\; ;a})^2 \, . \label{optical scalars}
\end{equation}
Purely geometric definition of the Kundt family of spacetimes, namely that it admits nonexpanding (${\Theta=0}$), nontwisting (${A=0}$) and shearfree~(${\sigma=0}$) such a congruence, implies that there exist suitable coordinates in which the line element of any Kundt spacetime can be written as~\cite{Kundt:1962,Stephani:2003,GriffithsPodolsky:2009,PodolskyZofka:2009,ColeyEtal:2009}
\begin{equation}
\dd s^2 = g_{ij}(u,x)\, \dd x^i\dd x^j+2\,g_{ui}(r,u,x)\,\dd x^i \dd u-2\,\dd u\,\dd r+g_{uu}(r,u,x)\,\dd u^2 \, . \label{obecny Kundtuv prostorocas}
\end{equation}
The coordinate $r$ is the affine parameter along the congruence ${k^a=\partial_r}$, ${u=\ }$const. are null (wave)surfaces, and ${x\equiv(x^2, x^3, \ldots, x^{D-1})}$ are ${D-2}$ spatial coordinates in the transverse Riemannian space. Notice that the spatial part $g_{ij}$ of the metric must be independent of $r$, all other metric components $g_{ui}$ and $g_{uu}$ can, in principle, be functions of all the coordinates $(r,u,x)$. No specific Einstein field equations have been employed yet.

For such most general Kundt line element  (\ref{obecny Kundtuv prostorocas}) a lengthy calculation gives the following components of the Riemann curvature tensor
\begin{eqnarray}
R_{rprq} \rovno 0\, , \label{Riemann rprq - Kundt} \\
R_{rpru} \rovno -\textstyle{\frac{1}{2}}g_{up,rr} \, , \label{Riemann rpru - Kundt} \\
R_{ruru} \rovno -\textstyle{\frac{1}{2}}g_{uu,rr}+\textstyle{\frac{1}{4}}g^{ij}g_{ui,r}g_{uj,r} \, , \label{Riemann ruru - Kundt} \\
R_{rpkq} \rovno 0 \, , \label{Riemann rpkq - Kundt} \\
R_{rpuq} \rovno \textstyle{\frac{1}{2}}g_{up,rq}+\textstyle{\frac{1}{4}}g_{up,r}g_{uq,r}-\textstyle{\frac{1}{4}}g^{ij}g_{ui,r}\left(2g_{j(p,q)}-g_{pq,j}\right) \, , \\
R_{rupq} \rovno g_{u[p,q],r} \, , \\
R_{ruup} \rovno g_{u[u,p],r}+\textstyle{\frac{1}{4}}g^{ri}g_{up,r}g_{ui,r}-\textstyle{\frac{1}{4}}g^{ij}g_{ui,r}\left(2g_{j(u,p)}-g_{up,j}\right) \, ,  \\
R_{kplq} \rovno \,^{S}R_{kplq} \, , \\
R_{upkq} \rovno g_{p[k,q],u}-g_{u[k,q],p} \nonumber \\
 &&\ +\textstyle{\frac{1}{4}}\left[g_{uk,r}\left(g_{pq,u}-2g_{u(p,q)}\right)-g_{uq,r}\left(g_{pk,u}-2g_{u(p,k)}\right)\right] \nonumber \\
 &&\ +\textstyle{\frac{1}{4}}g^{ri}\left[g_{uk,r}\left(2g_{i(p,q)}-g_{pq,i}\right)-g_{uq,r}\left(2g_{i(p,k)}-g_{pk,i}\right)\right] \nonumber \\
 &&\ +\textstyle{\frac{1}{4}}g^{ij}\left(2g_{j(u,q)}-g_{uq,j}\right)\left(2g_{i(p,k)}-g_{pk,i}\right) \nonumber \\
 &&\ -\textstyle{\frac{1}{4}}g^{ij}\left(2g_{j(u,k)}-g_{uk,j}\right)\left(2g_{i(p,q)}-g_{pq,i}\right) \, , \\
R_{upuq} \rovno g_{u(p,q),u}-\textstyle{\frac{1}{2}}\left(g_{pq,uu}+g_{uu,pq}\right) +\textstyle{\frac{1}{4}}g^{rr}g_{up,r}g_{uq,r}\nonumber \\
 &&\ -\textstyle{\frac{1}{4}}g_{uu,r}\left[2g_{u(p,q)}-g_{pq,u}-g^{ri}\left(2g_{i(p,q)}-g_{pq,i}\right)\right] \nonumber \\
 &&\ +\textstyle{\frac{1}{4}}g_{up,r}\left[g_{uu,q}-g^{ri}\left(2g_{i(u,q)}-g_{uq,i}\right) \right] \nonumber \\
 &&\ +\textstyle{\frac{1}{4}}g_{uq,r}\left[g_{uu,p}-g^{ri}\left(2g_{i(u,p)}-g_{up,i}\right)\right] \nonumber \\
 &&\ +\textstyle{\frac{1}{4}}g^{ij}\left(2g_{j(u,p)}-g_{up,j}\right)\left(2g_{i(u,q)}-g_{uq,i}\right) \nonumber \\
 &&\ -\textstyle{\frac{1}{4}}g^{ij}\left(2g_{uj,u}-g_{uu,j}\right)\left(2g_{i(p,q)}-g_{pq,i}\right)  \, , \label{Riemann upuq - Kundt}
\end{eqnarray}
where $i,j,k,l,p,q$ denote the spatial components (and derivatives w.r.t.)~$x$. The superscript ``$\,^{S}$'' labels tensor quantities corresponding to the spatial metric $g_{ij}$, with derivatives taken only with respect to the coordinates~$x$. The components of the Ricci tensor are
\begin{eqnarray}
R_{rr} \rovno 0\, , \label{Ricci rr - Kundt}\\
R_{rk}\rovno -\textstyle{\frac{1}{2}}g_{uk,rr}  \, ,\label{Ricci rk - Kundt}\\
R_{ru}\rovno -\textstyle{\frac{1}{2}}g_{uu,rr}+\textstyle{\frac{1}{2}}g^{ri}g_{ui,rr}+\textstyle{\frac{1}{2}}g^{pq}g_{up,rq} \nonumber \\
 &&\ +\textstyle{\frac{1}{2}}g^{pq}g_{up,r}g_{uq,r}-\textstyle{\frac{1}{4}}g^{pq}g^{ij}g_{ui,r}\left(2g_{jp,q}-g_{pq,j}\right) \, , \label{Ricci ru - Kundt} \\
 R_{pq}\rovno \!^{S}R_{pq}-g_{u(p,q),r}-\textstyle{\frac{1}{2}}g_{up,r}g_{uq,r}+\textstyle{\frac{1}{2}}g^{kl}g_{uk,r}\left(2g_{l(p,q)}-g_{pq,l}\right)\!, \label{Ricci pq - Kundt}\\
R_{uu}\rovno -\textstyle{\frac{1}{2}}g^{rr}g_{uu,rr}-2g^{ri}g_{u[u,i],r}+\textstyle{\frac{1}{2}}g^{pq}\left(2g_{up,uq}-g_{pq,uu}-g_{uu,pq}\right) \nonumber \\
 &&\ -\textstyle{\frac{1}{2}}g^{rp}g^{rq}g_{up,r}g_{uq,r}+\textstyle{\frac{1}{2}}g^{rr}g^{pq}g_{up,r}g_{uq,r} \nonumber \\
 &&\ +\textstyle{\frac{1}{2}}g^{pq}g^{ri}g_{up,r}(2g_{q(u,i)}-g_{ui,q}) \nonumber \\
 &&\ -\textstyle{\frac{1}{4}}g^{pq}g_{uu,r}\left[2g_{up,q}-g_{pq,u}-g^{ri}\left(2g_{ip,q}-g_{pq,i}\right)\right] \nonumber \\
 &&\ +\textstyle{\frac{1}{2}}g^{pq}g_{up,r}\left[g_{uu,q}-g^{ri}\left(2g_{i(u,q)}-g_{uq,i}\right) \right] \nonumber \\
 &&\ + \textstyle{\frac{1}{4}}g^{pq}g^{ij}\left(2g_{j(u,p)}-g_{up,j}\right)\left(2g_{i(u,q)}-g_{uq,i}\right)\nonumber \\
 &&\ - \textstyle{\frac{1}{4}}g^{pq}g^{ij}\left(2g_{uj,u}-g_{uu,j}\right)\left(2g_{ip,q}-g_{pq,i}\right) \, , \label{Ricci uu - Kundt}\\
R_{uk}\rovno -\textstyle{\frac{1}{2}}g^{rr}g_{uk,rr}-g_{u[u,k],r}+g^{ri}(g_{u[i,k],r}-g_{k[u,i],r}) \nonumber \\
 &&\ +g^{pq}(g_{p[k,q],u}-g_{u[k,q],p})-\textstyle{\frac{1}{2}}g^{ri}g_{uk,r}g_{ui,r} \nonumber \\
 &&\ +\textstyle{\frac{1}{4}}g^{pq}g^{ri}\left[4g_{uq,r}g_{k[p,i]}+g_{uk,r}(2g_{i(p,q)}-g_{pq,i})\right] \nonumber \\
 &&\ +\textstyle{\frac{1}{4}}g^{pq}\left[2g_{up,r}g_{uq,k}-g_{uk,r}(2g_{up,q}-g_{pq,u})\right] \nonumber \\
 &&\ +\textstyle{\frac{1}{4}}g^{pq}g^{ij}\left(2g_{j(u,q)}-g_{uq,j}\right)\left(2g_{i(p,k)}-g_{pk,i}\right) \nonumber \\
 &&\ -\textstyle{\frac{1}{4}}g^{pq}g^{ij}\left(2g_{j(u,k)}-g_{uk,j}\right)\left(2g_{ip,q}-g_{pq,i}\right)\, , \label{Ricci uk - Kundt}
\end{eqnarray}
and the Ricci scalar curvature of the Kundt spacetime (\ref{obecny Kundtuv prostorocas}) is given by
\begin{eqnarray}
R \rovno \,^{S}R+g_{uu,rr}-2g^{ri}g_{ui,rr}-2g^{pq}g_{up,rq} \nonumber \\
 &&\ -{\textstyle\frac{3}{2}}g^{pq}g_{up,r}g_{uq,r}+g^{pq}g^{kl}g_{uk,r}(2g_{lp,q}-g_{pq,l})\, . \label{Ricci scalar - Kundt}
\end{eqnarray}

\section{Applying the field equations}

So far we have not specified the matter content of the spacetimes. Now, following the approach presented in~\cite{PodolskyZofka:2009}, we can  determine the $r$-dependence of the metric (\ref{obecny Kundtuv prostorocas}) using the Einstein field equations ${R_{ab}-\frac{1}{2}R\,g_{ab}+\Lambda g_{ab}=8\pi T_{ab}}$. Since ${R_{rr}=0}$ and ${g_{rr}=0}$, there is an obvious restriction on the energy-momentum tensor allowed in the Kundt family, namely ${T_{rr}=0}$. \emph{Assuming} ${T_{rk}=0}$, we can  directly integrate the Einstein equation $R_{rk}=0$ with (\ref{Ricci rk - Kundt}), yielding $g_{uk}$ \emph{linear} in~$r$. Using the field equation ${R_{ru}+\frac{1}{2}R-\Lambda=8\pi T_{ru}}$, this implies that the component $T_{ru}$ must be independent of $r$. Taking the trace of Einstein's equations we can also determine the $r$-dependence of $g_{uu}$: if the trace $T$ of energy-momentum tensor $T_{ab}$ does not depend on the coordinate~$r$, the metric function $g_{uu}$ can only be (at most) \emph{quadratic} in~$r$, see (\ref{Ricci scalar - Kundt}). Under these conditions
\begin{equation}
\dd s^2 = g_{ij} \,\dd x^i\dd x^j+2\,(e_i+ f_i \,r)\,\dd x^i \dd u-2\,\dd u\,\dd r+(a\,r^2+b\,r+c)\,\dd u^2 \, , \label{obecny Kundt po uziti ER}
\end{equation}
where all the functions ${g_{ij}, e_i, f_i, a, b}$ and $c$ are independent of $r$, and are constrained by the specific Einstein equations~\cite{PodolskyZofka:2009}. In particular, any \emph{vacuum} Kundt metric, possibly with a cosmological constant $\Lambda$ and/or aligned electromagnetic field, can be written in the form (\ref{obecny Kundt po uziti ER}).

\section{Geodesic deviation in an arbitrary spacetime}

\begin{figure}
  \begin{center}
   \vspace{-1mm}
  \includegraphics[width=0.6\textwidth]{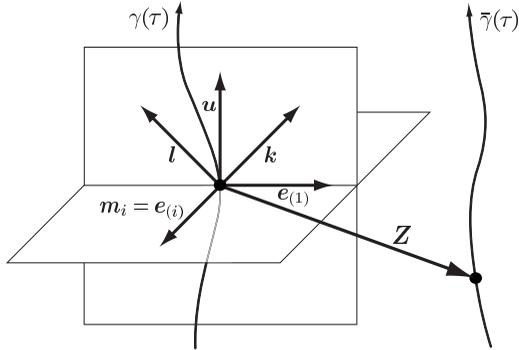}
  \vspace{-6mm}
  \end{center}
  \caption{\label{figure1} Evolution of the separation vector $\boldZ$ that connects particles moving along  geodesics $\gamma(\tau)$, $\bar\gamma(\tau)$ is given by the equation of geodesic deviation (\ref{invariant form of eq of geodesic deviationL}) and (\ref{invariant form of eq of geodesic deviationT}). Its components are expressed in the orthonormal frame ${\{\bolde_a \}}$,  ${\bolde_{(0)}=\boldu}$. The associated null frame ${\{\boldk, \boldl, \boldm_{i} \}}$ is also indicated.}
\end{figure}

In our recent work \cite{PodolskySvarc:2012} we demonstrated that the equation of geodesic deviation, which describes relative motion of nearby free test particles, can in \emph{any} $D$-dimensional spacetime  be expressed in the invariant form
\begin{eqnarray}
&&\hspace{-14mm}\ddot{Z}^{(1)} = \frac{2\Lambda}{(D-2)(D-1)}\,Z^{(1)} +\Psi_{2S}\,Z^{(1)}
+ \frac{1}{\sqrt{2}}\,(\,\Psi_{1T^j}-\Psi_{3T^j})\,Z^{(j)} \nonumber \\
 && \hspace{-2.5mm} + \, \frac{8\pi}{D-2}\!\left[T_{(1)(1)}\,Z^{(1)}+T_{(1)(j)}\,Z^{(j)}-\Big(T_{(0)(0)}+\frac{2\,T}{D-1}\Big)Z^{(1)}\right]\!, \label{invariant form of eq of geodesic deviationL}\\
&&\hspace{-13.8mm}\ddot{Z}^{(i)} = \frac{2\Lambda}{(D-2)(D-1)}\,Z^{(i)}-\Psi_{2T^{(ij)}}\,Z^{(j)}
+ \frac{1}{\sqrt{2}}\,(\,\Psi_{1T^i}-\Psi_{3T^i})\,Z^{(1)}\nonumber \\
 && \hspace{+55.8mm} -\frac{1}{2}\,(\,\Psi_{0^{ij}}+\Psi_{4^{ij}})\,Z^{(j)}\nonumber\\
 && \hspace{-2.5mm} + \, \frac{8\pi}{D-2}\!\left[T_{(i)(1)}\,Z^{(1)}+T_{(i)(j)}\,Z^{(j)}-\Big(T_{(0)(0)}+\frac{2\,T}{D-1}\Big)Z^{(i)}\right]\!, \label{invariant form of eq of geodesic deviationT}
\end{eqnarray}
with ${\,i,j=2,\ldots,D-1\,}$. Here ${Z^{(1)}, Z^{(2)}, \ldots, Z^{(D-1)}}$ are spatial components of the separation vector ${\boldZ=Z^a \,\bolde_a}$ between the test particles in a natural interpretation orthonormal frame ${\{\bolde_a \}}$ where ${\bolde_{(0)}=\boldu}$ is the velocity vector of the fiducial test particle (${\bolde_a \cdot \bolde_b=\eta_{ab}}$),  ${\ddot{Z}^{(1)}, \ddot{Z}^{(2)}, \ldots, \ddot{Z}^{(D-1)}}$ are the corresponding relative accelerations, $T_{ab}$ are frame components of the energy-momentum tensor, and the scalars $\Psi_{A^{...}}$ defined as
\begin{eqnarray}
\Psi_{0^{ij}} \rovno C_{abcd}\; k^a\, m_i^b\, k^c\, m_j^d \, , \nonumber \\
\Psi_{1T^{i}} \rovno C_{abcd}\; k^a\, l^b\, k^c\, m_i^d \, ,
\hspace{10mm} \Psi_{1^{ijk}} = C_{abcd}\; k^a\, m_i^b\, m_j^c\, m_k^d    \ ,\nonumber \\
\Psi_{2S} \rovno C_{abcd}\; k^a\, l^b\, l^c\, k^d \, ,
\hspace{11mm} \Psi_{2^{ijkl}}= C_{abcd}\; m_i^a\, m_j^b\, m_k^c\, m_l^d \, ,\nonumber \\
\Psi_{2T^{ij}}\rovno C_{abcd}\; k^a\, m_i^b\, l^c\, m_j^d \, ,
\hspace{10.2mm} \Psi_{2^{ij}}  = C_{abcd}\; k^a\, l^b\, m_i^c\, m_j^d \, ,\nonumber \\
\Psi_{3T^{i}} \rovno C_{abcd}\; l^a\, k^b\, l^c\, m_i^d \, ,
\hspace{10.9mm} \Psi_{3^{ijk}} = C_{abcd}\; l^a\, m_i^b\, m_j^c\, m_k^d \, ,\nonumber\\
\Psi_{4^{ij}} \rovno C_{abcd}\; l^a\, m_i^b\, l^c\, m_j^d \, , \label{defPsiCoef}
\end{eqnarray}
${\,i,j,k,l=2,\ldots,D-1\,}$, are components of the Weyl tensor with respect to the null frame ${\{\boldk, \boldl, \boldm_{i} \}}$ associated with ${\{\bolde_a \}}$ via the relations ${\boldk=\ssqrt(\boldu+\bolde_{(1)})}$, ${\boldl=\ssqrt(\boldu-\bolde_{(1)})}$, ${\boldm_{i}=\bolde_{(i)}}$, see figure~\ref{figure1}.

Components of the Weyl tensor (\ref{defPsiCoef}) are listed by their boost weight and directly generalize the standard Newman--Penrose complex scalars $\Psi_A$ known from the ${D=4}$ case~\cite{KrtousPodolsky:2006,PodolskySvarc:2012}. In equations (\ref{invariant form of eq of geodesic deviationL}), (\ref{invariant form of eq of geodesic deviationT}), only the ``electric part'' of the Weyl tensor represented by the scalars in the left column of (\ref{defPsiCoef}) occurs, and there are various constraints and symmetries, for example
\begin{eqnarray}
&&\Psi_{1T^i} = \Psi_{1^{k}}{}^{_k}{}_{^i} \, ,\
\Psi_{2S}   = {\textstyle\frac{1}{2}}\Psi_{2^{kl}}{}^{_{kl}} \, ,\ \Psi_{2T^{(ij)}}={\textstyle\frac{1}{2}}\Psi_{2^{ikj}}{}^{_k}\, ,\
\Psi_{3T^i} = \Psi_{3^{k}}{}^{_k}{}_{^i} \, ,\nonumber \\
&&\Psi_{0^{ij}} = \Psi_{0^{(ij)}} \, ,\hspace{2.4mm}  \Psi_{0^{k}}{}^{_k} = 0 \, ,\hspace{9.85mm}
\Psi_{4^{ij}} = \Psi_{4^{(ij)}} \, ,\hspace{2.4mm}  \Psi_{4^{k}}{}^{_k} = 0 \, .\label{symmetries}
\end{eqnarray}
\vspace{-10mm}

\section{Geodesic deviation in Kundt spacetimes}

For the general Kundt spacetime (\ref{obecny Kundtuv prostorocas}), the null interpretation frame adapted to an arbitrary observer moving along a timelike geodesic $\gamma(\tau)$ with velocity $\boldu=\dot{r}\,\mathbf{\partial}_{r}+\dot{u}\,\mathbf{\partial}_{u}+\dot{x}^i\mathbf{\partial}_{i}$ takes the form
\begin{eqnarray}
\boldk \rovno \frac{1}{\sqrt{2}\,\dot{u}}\,\mathbf{\partial}_r \ , \nonumber \\
\boldl\rovno \Big(\sqrt{2}\,\dot{r}-\frac{1}{\sqrt{2}\,\dot{u}}\Big)\mathbf{\partial}_r+\sqrt{2}\,\dot{u}\,\mathbf{\partial}_u+\sqrt{2}\,\dot{x}^i\mathbf{\partial}_i \, , \nonumber \\
\boldm_i\rovno \frac{1}{\dot{u}}\,m_{i}^{j}(g_{ju}\,\dot{u}+g_{jk}\,\dot{x}^k)\,{\partial}_r+m_i^j\,\mathbf{\partial}_j \, , \label{Kundt interpretation frame}
\end{eqnarray}
where $m_i^j$ satisfy ${{\,g_{jl}\,m_i^j \,m_k^l}=\delta_{ik}}$ to fulfil ${\boldm_i \cdot \boldm_k=\delta_{ik}}$,  ${\boldk \cdot \boldl=-1}$. Vector $\boldk$ is oriented along the nonexpanding, nontwisting and shearfree null congruence ${k^a=\partial_r}$ defining the Kundt family. Moreover, ${\boldu=\ssqrt{(\boldk+\boldl)}}$ and ${\bolde_{(1)}=\ssqrt{(\boldk-\boldl)}=\sqrt{2}\,\boldk-\boldu}$. The spatial vector ${\bolde_{(1)}}$ is thus uniquely determined by the geometrically privileged null congruence of the Kundt family, and the observer's velocity $\boldu$. For this reason we call such a special direction ${\bolde_{(1)}}$ \emph{longitudinal}, while the ${D-2}$ directions ${\bolde_{(i)}=\boldm_{i}}$ \emph{transverse}.

In order to evaluate the scalars (\ref{defPsiCoef}) we need to calculate the Weyl tensor
\begin{equation}\label{Decomp}
C_{abcd}=R_{abcd}-\frac{2}{D-2}(\,g_{a[c}\,R_{d]b}-g_{b[c}\,R_{d]a})+\frac{2\,R\,g_{a[c}\,g_{d]b}}{(D-1)(D-2)}\, ,
\end{equation}
using the components of the Riemann and Ricci tensors (\ref{Riemann rprq - Kundt})--(\ref{Ricci scalar - Kundt}).
We immediately observe that ${C_{rprq}=0}$ which implies ${\Psi_{0^{ij}}=0}$. Therefore, \emph{all Kundt spacetimes are of algebraic type} $\mathrm{I}$, or more special, and $\partial_r$ is~WAND.

Restricting now to the important subfamily (\ref{obecny Kundt po uziti ER}) for which
\begin{equation}\label{specgui}
g_{ui}=e_i(u,x)+ f_i(u,x) \,r\, ,\qquad
g_{uu}=a(u,x)\,r^2+b(u,x)\,r+c(u,x)\, ,
\end{equation}
we obtain ${R_{rpru}=0}$, ${R_{rp}=0}$ which implies ${C_{rpru}=0}$, ${C_{rpkq}=0}$ so that ${\Psi_{1T^{i}}=0}$, ${\Psi_{1^{ijk}}=0}$.
Since all Weyl scalars of boost weights $2$ and~$1$ vanish, the metric (\ref{obecny Kundt po uziti ER}) represents Kundt spacetimes of algebraic type~$\mathrm{II}$ (or more special). Equations (\ref{invariant form of eq of geodesic deviationL}), (\ref{invariant form of eq of geodesic deviationT}) for the geodesic deviation (omitting the frame components of $T_{ab}$ encoding the direct influence of matter) in the case of the Kundt class of spacetimes (\ref{obecny Kundt po uziti ER}) thus simplify considerably to
\begin{eqnarray}
\ddot{Z}^{(1)} \rovno \frac{2\Lambda}{(D-2)(D-1)}\,Z^{(1)} +\Psi_{2S}\,Z^{(1)}
- \frac{1}{\sqrt{2}}\,\Psi_{3T^j}\,Z^{(j)} \, , \label{Kundt geodesic deviation}\\
\ddot{Z}^{(i)} \rovno \frac{2\Lambda}{(D-2)(D-1)}\,Z^{(i)}-\Psi_{2T^{(ij)}}\,Z^{(j)}
- \frac{1}{\sqrt{2}}\,\Psi_{3T^i}\,Z^{(1)} -\frac{1}{2}\,\Psi_{4^{ij}}\,Z^{(j)}\, ,\nonumber
\end{eqnarray}
where the only nonvanishing Weyl scalars are
\begin{eqnarray}
\Psi_{2S}\rovno -R_{ruru}+\textstyle{\frac{2}{D-2}}R_{ru}+\textstyle{\frac{1}{(D-1)(D-2)}}R \, , \nonumber \\
\Psi_{2T^{ij}} \rovno m_{i}^{p}m_{j}^{q}\textstyle{\big[R_{rpuq}-\frac{1}{D-2}\left(g_{pq}R_{ru}-R_{pq}\right)-\frac{1}{(D-1)(D-2)}R\,g_{pq}\big]} \, , \nonumber \\
\Psi_{3T^j} \rovno \sqrt{2}\,m_j^{p}\Big\{\dot{x}^k\big[R_{ruru}\,g_{kp}-R_{rkup}-R_{rukp}-\textstyle{\frac{1}{D-2}}\left(g_{kp}R_{ru}+R_{kp}\right)\big] \nonumber \\
 &&\qquad\ +\dot{u}\,\big[R_{ruru}\,g_{up}-R_{ruup}-\textstyle{\frac{1}{D-2}}\left(g_{up}R_{ru}+R_{up}\right)\big]\Big\} \, , \nonumber \\
\Psi_{4^{ij}} \rovno 2\,m_{(i}^{p}m_{j)}^{q}\Big\{
\dot{x}^k\dot{x}^l\big[R_{rpuq}\,g_{kl}-g_{pk}(2R_{rluq}-g_{lq}R_{ruru}+2R_{rulq}) \nonumber \\
 && \hspace{19.0mm} +R_{kplq}-\textstyle{\frac{1}{D-2}}g_{pq}\left(g_{kl}R_{ru}+R_{kl}\right)\big] \nonumber \\
 && \hspace{10.0mm} +2\dot{u}\dot{x}^k\big[R_{rpuq}\,g_{uk}-g_{up}(R_{rkuq}-R_{ruru}\,g_{qk}+R_{rukq}) \nonumber \\
 && \hspace{19.0mm} -R_{ruup}\,g_{qk}+R_{upkq}-\textstyle{\frac{1}{D-2}}g_{pq}(g_{uk}R_{ru}+R_{uk})\big] \nonumber \\
 && \hspace{10.0mm} +\dot{u}^2\big[R_{rpuq}\,g_{uu}-g_{uq}\left(2R_{ruup}-g_{up}R_{ruru}\right)+R_{upuq} \nonumber \\
 && \hspace{19.0mm} -\textstyle{\frac{1}{D-2}}g_{pq}\left(g_{uu}R_{ru}+R_{uu}\right)\big]\Big\} \, , \label{Psi - obecny Kundt - rozepsane}
\end{eqnarray}
and the components $R_{abcd}$ are explicitly given by (\ref{Riemann ruru - Kundt})--(\ref{Riemann upuq - Kundt}), $R_{ab}$ by (\ref{Ricci ru - Kundt})--(\ref{Ricci uk - Kundt}), and the Ricci scalar curvature $R$ is given by~(\ref{Ricci scalar - Kundt}).

The relative motion of free test particles in any Kundt spacetime (\ref{obecny Kundt po uziti ER}) is thus composed of the \emph{isotropic influence} of the cosmological constant~$\Lambda$, \emph{Newton-like tidal deformations} represented by $\Psi_{2S}$, $\Psi_{2T^{(ij)}}$, \emph{longitudinal} accelerations associated with the direction $+\mathbf{e}_{(1)}$ given by $\Psi_{3T^j}$, and by \emph{transverse gravitational waves} propagating along $+\mathbf{e}_{(1)}$ encoded in the symmetric traceless matrix $\Psi_{4^{ij}}$, see (\ref{symmetries}). The invariant amplitudes (\ref{Psi - obecny Kundt - rozepsane}) combine the curvature of the Kundt spacetime with kinematics of the specific geodesic motion. In contrast to longitudinal and transverse wave effects, the Newton-like deformations caused by $\Psi_{2S}$ and $\Psi_{2T^{(ij)}}$ are independent of the observer's velocity components $\dot{x}^i$ and $\dot{u}$.

\section{Discussion of particular subfamilies}

The Kundt class involves several physically interesting subfamilies, for example pp-waves including gyratons and VSI spacetimes.

The \emph{pp-waves} are defined by admitting a covariantly constant null vector field $k^a$~\cite{Stephani:2003,GriffithsPodolsky:2009}. They thus belong to the Kundt class with all metric functions independent of $r$, which is the metric (\ref{obecny Kundt po uziti ER}) with ${ f_i=0, a=0=b}$:
\begin{equation}
\dd s^2 = g_{ij}(u,x) \,\dd x^i\dd x^j+2\,e_i(u,x)\,\dd x^i \dd u-2\,\dd u\,\dd r+c(u,x)\,\dd u^2 \, . \label{pp}
\end{equation}
The components $e_i$ encode the possible presence of gyratonic matter.

The \emph{VSI spacetimes} have the property that their scalar curvature invariants of all orders vanish identically. As shown in \cite{ColeyEtal:2006}, these spacetimes must be of the form (\ref{obecny Kundt po uziti ER}) with  flat transverse space ${g_{ij}=\delta_{ij}}$:
\begin{equation}
\dd s^2 = \delta_{ij} \,\dd x^i\dd x^j+2\,(e_i+ f_i \,r)\,\dd x^i \dd u-2\,\dd u\,\dd r+(a\,r^2+b\,r+c)\,\dd u^2 \, . \label{VSI}
\end{equation}

It is straightforward to apply our general results (\ref{Kundt geodesic deviation}) to these particular subcases by evaluating the corresponding Weyl scalars (\ref{Psi - obecny Kundt - rozepsane}) and discussing their specific influence on test particles. We have to restrict ourselves only to the simplest case here,\footnote{A thorough discussion of other cases will follow in our subsequent paper.} to \emph{vacuum VSI pp-waves without gyratons}:
\begin{equation}
\dd s^2 = \delta_{ij}\,\dd x^i\dd x^j-2\,\dd u\,\dd r+c(u,x)\,\dd u^2 \, . \label{ppsimplest}
\end{equation}
Since ${\Lambda=0}$, ${\Psi_{2S}=0=\Psi_{2T^{ij}}}$, ${\Psi_{3T^j}=0}$, the geodesic deviation reduces~to \begin{equation}
\ddot{Z}^{(1)}=0 \, ,\qquad
\ddot{Z}^{(i)}=-{\textstyle\frac{1}{2}}\,\Psi_{4^{ij}}\,Z^{(j)}\, .\label{pp geodesic deviation}
\end{equation}
This clearly represents \emph{gravitational waves} propagating along the spatial direction ${(1)}$, with the test particles influenced only in the \emph{transverse} directions ${(i)=(2), (3), \ldots, (D-1)}$. The elements of the \emph{symmetric~and traceless} ${(D-2)\times(D-2)}$ matrix ${\Psi_{4^{ij}}=-\dot{u}^2\,c_{,ij}}$ (where $\dot{u}$ is a constant) directly encode the corresponding wave amplitudes. Obviously, there are ${\frac{1}{2}D(D-3)}$ independent \emph{polarization states}.

Taking, e.g., a quadratic function
${c(x)\equiv\sum_{i=2}^{D-1} {\cal A}_i\, (x^i)^2}$ where the constants must satisfy ${\sum_{i=2}^{D-1} {\cal A}_i=0}$, the wave-amplitude matrix becomes ${\Psi_{4^{ij}}=-2\dot{u}^2\, \hbox{diag}({\cal A}_2,{\cal A}_3,\ldots)}$. Relative motion of (initially static) particles given by (\ref{pp geodesic deviation}) can  be explicitly integrated: in the spatial directions with \emph{positive eigenvalues} ${{\cal A}_i>0}$ they \emph{recede} as ${Z^{(i)}(\tau)=Z^{(i)}_0\cosh\left(\sqrt{{\cal A}_i}\,|\dot u|\,\tau   \right) }$, while with \emph{negative eigenvalues} ${{\cal A}_i<0}$ they \emph{converge} as ${Z^{(i)}(\tau)=Z^{(i)}_0\cos\left(\sqrt{{\cal A}_i}\,|\dot u|\,\tau   \right) }$, and in the directions where ${{\cal A}_i=0}$ the particles stay fixed, ${Z^{(i)}(\tau)=Z^{(i)}_0}$.

\begin{figure}[h]
  \begin{center}
  \includegraphics[width=0.25\textwidth]{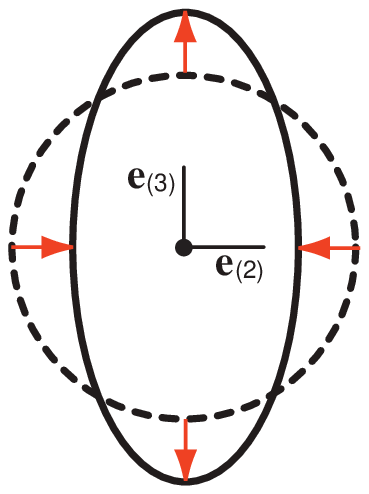}
  \hspace{15mm}
  \includegraphics[width=0.25\textwidth]{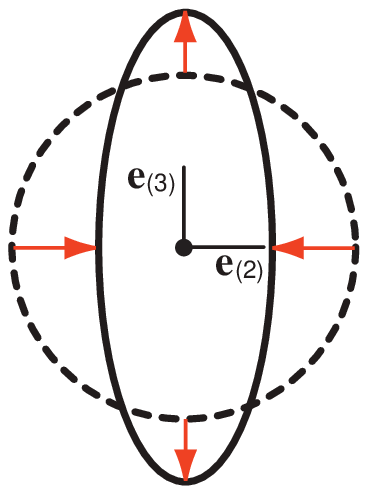}
  \end{center}
   \vspace{-3mm}
  \caption{\label{figure2} Standard (left) and one of peculiar deformations of a detector indicating extension of the gravitation wave into higher dimensions (right).}
\end{figure}

\nopagebreak
In principle, the presence of higher-dimensional components of gravitational waves could be observed by detectors in our (1+3)-dimensional universe as the \emph{violation of the standard TT-property}. Indeed, taking the simplest case ${D=5}$, the matrix reads ${\Psi_{4^{ij}}=-2\dot{u}^2\, \hbox{diag}({\cal A}_2,{\cal A}_3,{\cal A}_4)}$ where ${{\cal A}_2=-({\cal A}_3+{\cal A}_4)}$. In the absence of the higher-dimensional component,  ${{\cal A}_4=0}$, an interferometer in our space detects usual deformations shown in the left part of figure~\ref{figure2}. But if ${{\cal A}_4\not=0}$ then ${{\cal A}_2\not=-{\cal A}_3}$, and a peculiar deformation, such as on the right part of figure~\ref{figure2}, would be observed.

   \vspace{-5mm}

\section*{Acknowledgement}

R.\v{S}. was supported by the grants GA\v{C}R 202/09/0772 and SVV-265301, and J.P. by the grants GA\v{C}R P203/12/0118 and MSM0021620860.

\section*{References}
\bibliography{ae100prg}

\end{document}